\documentclass[useAMS,usenatbib,usegraphicx]{mn2e}

\usepackage{times}
\usepackage{amssymb}
\usepackage{graphics,epsfig}

\def\arcs{\mbox{\ensuremath{^{\prime\prime}}}}
\def\farcs{\mbox{\ensuremath{.\!\!\arcs}}}

\def\hi{H~{\sc i}}
\def\nhi{N(H\,{\sc i})}
\def\sii{Si~{\sc ii}}
\def\mgii{Mg~{\sc ii}}
\def\feii{Fe~{\sc ii}}
\newcommand{\kms}{km~s$^{-1}$}
\newcommand{\cm}{cm$^{-2}$}

\newcommand{\ts}{{T_s}}
\newcommand{\beq}{\begin{equation}}
\newcommand{\eeq}{\end{equation}}

\title[A low spin temperature DLA at $z = 2.289$]{Discovery of 21cm absorption in a $z_{\rm abs} =2.289$ DLA towards TXS~0311+430: The first low spin temperature absorber at $z>1$.}
\author[B. York et al.]{Brian A. York$^{1}$\thanks{E-mail: briany@uvic.ca (BY), 
nkanekar@aoc.nrao.edu (NK), sarae@uvic.ca (SLE), pettini@ast.cam.ac.uk (MP)}, 
Nissim Kanekar$^{2}$\footnotemark[1], 
Sara L. Ellison$^{1}$\footnotemark[1] and
Max Pettini$^{3}$\footnotemark[1]\\
$^{1}$Department of Physics and Astronomy, University of Victoria, 3800 Finnerty 
    Road, Victoria, BC V8P 1A1, Canada\\
$^{2}$National Radio Astronomy Observatory, 1003 Lopezville Road, Socorro, NM 
    87801, USA\\
$^{3}$Institute of Astronomy, University of Cambridge, Madingley Road, Cambridge 
    CB3 0HA}
\begin{document}

\date{}

\pagerange{\pageref{firstpage}--\pageref{lastpage}} \pubyear{2007}

\maketitle

\label{firstpage}

\begin{abstract}
We report the detection of \hi~21~cm absorption from the $z=2.289$
damped Lyman-$\alpha$ system (DLA) towards TXS~0311+430, with the
Green Bank Telescope.  The 21~cm absorption has a velocity spread
(between nulls) of $\sim 110$~km~s$^{-1}$ and an integrated optical
depth of $\int \tau {\rm d}V = (0.818 \pm 0.085)$~km~s$^{-1}$.  We
also present new Giant Metrewave Radio Telescope 602~MHz imaging of
the radio continuum.  TXS~0311+430 is unresolved at this frequency,
indicating that the covering factor of the DLA is likely to be high.
Combining the integrated optical depth with the DLA \hi~column density
of \nhi\ = $(2 \pm 0.5) \times 10^{20}$~\cm, yields a spin temperature
of $\ts = (138 \pm 36)$~K, assuming a covering factor of unity.  This
is the first case of a low spin temperature ($< 350$~K) in a $z
> 1$ DLA and is among the lowest ever measured in any DLA. Indeed, 
the $\ts$ measured for this DLA is similar to values measured in the 
Milky Way and local disk galaxies. We also determine a lower
limit (Si/H)~$\gtrsim 1/3$ solar for the DLA metallicity, amongst the highest
abundances measured in DLAs at any redshift.  Based on low redshift
correlations, the low $\ts$, large 21~cm absorption width
and high metallicity all suggest that the $z \sim 2.289$ DLA is likely
to arise in a massive, luminous disk galaxy.
\end{abstract}


\section{Introduction}
\label{sec:intro}

Damped Lyman-$\alpha$ systems (DLAs) are the highest column density
absorbers seen along QSO lines-of-sight, with neutral hydrogen
column densities \nhi\ $\ge 2.0\times 10^{20}$~\cm. They have long been
identified as the precursors of today's galaxies and the primary gas
reservoir for star formation at high redshifts.  Despite the
recognised importance of the absorbers, their typical size, structure
and internal physical conditions remain issues of controversy
(e.g. \citealt{wolfe05}). For example, DLA metallicities, now measured
in over 100 absorbers, show very little evolution between $z
\sim 3$ and $z \sim 0$, with low-metallicity ($-2<$~[Z/H]~$< -1$,
where Z~$\equiv$~Zn, S or Si\footnote{In the usual notation,
[Z/H]~$\equiv$~log(N(Z)/N(\hi))$-$log(N(Z)/N (\hi))$_{\odot}$.  We use
solar values from \citet{lodders03}.}) absorbers the norm at all
redshifts (e.g. \citealt{kulkarni05}).  This lack of metallicity
evolution runs contrary to expectations that the interstellar
metallicity should rise towards lower redshifts if DLAs trace the bulk
of gas in galaxies.  Moreover, only a few tens of DLAs have their galactic 
counterparts identified, of which only a small fraction have been 
spectroscopically confirmed (e.g. \citealt{chen03}). Our understanding of 
the basic physical
properties of the absorbing galaxies, such as their mass, size, star
formation rates and luminosity, remains very limited.

\hi~21~cm absorption studies of DLAs towards radio-loud quasars
provide an independent probe of physical conditions in the
absorbers. They can be combined with optical measurements of the \hi\
column density of the DLA (from the Lyman-$\alpha$ line) to obtain the
column-density-weighted harmonic mean spin temperature $\ts$ of the
absorbing gas, allowing one to determine the temperature distribution
of the \hi\ along the line of sight.  Measurements of $\ts$ in a large 
number of DLAs may ultimately be used to infer other galactic properties 
of high redshift systems. For example, in the local Universe, large spiral 
disks like the Milky Way and M31 typically have low $\ts$ values ($\lesssim 300$~K; 
\citealt{braun92}) while high $\ts$ values ($\sim 1000$~K; \citealt{young97}) 
are more common in dwarf galaxies. Tentative evidence for a similar trend has
been found in low~$z$ DLAs, out to $z \sim 0.7$ \citep{kanekar03}.
If such correlations hold for DLAs at all redshifts, measurements of $\ts$, 
and its evolution, would provide interesting insights into galaxy evolution.

Unfortunately, despite a number of searches over the past two and a
half decades (e.g. \citealt{briggs83,carilli96,chengalur00,kanekar03}), spin
temperature estimates are available for only $15$~DLAs (of which
$10$ are lower limits) at $z \gtrsim 1.7$ (Kanekar et al, in preparation). 
All of these previous high $z$ measurements yield high spin temperatures 
of $\ts \gg 300$ K.  There are a number of reasons why the crop of $\ts$ 
estimates has increased slowly over the last 25 years, including observational 
issues such as the frequency coverage of radio telecopes and widespread radio
frequency interference (RFI) at the low frequencies ($\lesssim 1$~GHz)
of the redshifted 21~cm line. However, an additional important reason
for the relatively-small present 21~cm absorption sample is simply the
dearth of known DLAs towards radio-loud QSOs suitable for 21~cm 
absorption follow-up. We have hence been conducting an optical
survey of low-frequency-selected radio-loud quasars, specifically
designed to increase the number of $\ts$ estimates in the redshift
range $2<z<4$.  While the survey is still in progress, we report here
its first results, the detection of 21~cm absorption from the $z \sim
2.289$ DLA towards TXS~0311+430, the first case of a low spin temperature
in a high-$z$ DLA.

\section{Observations and data analysis}

\subsection{Optical observations}

Optical observations of $\sim$ 50~QSOs selected from the Texas 365~MHz 
survey \citep{douglas96} have been conducted at various facilities in order to
identify DLAs suitable for 21~cm follow-up.  As part of this
campaign, we observed TXS~0311+430 (B=21.5, $z_{\rm em} = 2.87$) with the
Gemini Multi-Object Spectrograph (GMOS) on the Gemini-North
telescope.  We obtained seven 2300-second and one 1700-second
long-slit spectra of TXS~0311+430, with a 1.0~arcsecond slit and the
B600\_G5303 disperser. The central wavelength was set to 4590\,\AA\ for four
of the exposures and to 4620\,\AA\ for the other four, so
as to achieve continuous wavelength coverage despite the
gap between CCD chips in the GMOS detector. The CCD
was binned $2 \times 2$. The final spectrum has a resolution
of 4.1\,\AA\ (full-width-at-half-maximum) and extends from $\sim 3550$ to 
6050\,\AA; this range allows the detection of DLAs in the redshift interval
$1.92 \leq z \leq 2.87$, with the upper limit set by the quasar
redshift.

The GMOS data were reduced using standard IRAF routines including bias
subtraction, flatfield correction, extraction using APALL, wavelength
fitting [the root-mean-square (RMS) error on the wavelength fits was
$\leq 0.14$~\AA], and finally, vacuum and heliocentric velocity
wavelength corrections; the full procedure for this and the other
optical spectra of our survey will be described in York et al. (in
preparation).  The final signal-to-noise ratio (S/N) ranged from $\sim
8$ per pixel at 3600~\AA\ to $\sim 20$ per pixel at 6000~\AA. The GMOS 
observations resulted in the detection of a DLA at $z \sim 2.289$ (see 
Section~\ref{sec:lya_metals}).

\subsection{Radio observations}
\label{sec:radio1}

An initial search for 21~cm absorption at the DLA redshift was carried out 
with the PF1-450~MHz receiver of the Green Bank Telescope
(GBT; program AGBT-06B-042) on September~12, 2006. We used
the GBT Spectral Processor as the backend, with a bandwidth of
1.25~MHz centred at 431.808~MHz, two circular polarizations, 1024
spectral channels and a spectral resolution of $\sim 0.85$~km/s
(before any smoothing).  The data were taken in total power mode, with
On/Off position-switching (with a 10-minute On/Off cycle made up of
2-second integrations) and online measurements of the system
temperature using a noise diode. The total on-source time was $\sim
35$~minutes.

Weak absorption was detected at the expected redshifted 21cm frequency
($\sim 431.8$~MHz) in the September run. We hence repeated the
observations on October~20, 2006, and January~4, 2007, to confirm the
feature. The same observational setup was used in these observing
sessions, except for the use of linear instead of circular
polarizations (as laboratory calibration information was not available
for the circular polarizations; we will hence not further discuss the
September data). The total on-source time on TXS~0311+430 was $\sim 50$~minutes 
and $\sim 65$~minutes in October and January, respectively. A calibrator, 
PKS~B0316+162, was also observed during the October session, with the same setup, 
to test for RFI at the observing frequency. The calibrator was observed for a 
total of 25~on-source minutes, broken into two runs, alternating with two 
runs on TXS~0311+430.

The GBT data were analyzed in AIPS++ using the \texttt{dish} package
of single-dish routines. After the initial data-editing, to remove
scans with correlator problems and RFI, the data were calibrated
(assuming a telescope gain of 2~K/Jy) and averaged together to measure
the flux density of TXS~0311+430. This yielded flux densities of $\sim
(6.2 \pm 0.7)$~Jy in October and $\sim (6.7 \pm 0.7)$~Jy in January, where
the errors include those from confusing sources in the primary beam
(note that \citealt{ficarra85} measured $(4.94 \pm 0.10)$~Jy at
408~MHz). A second-order spectral baseline was then fit to RFI- and
line-free channels for each 2-second spectrum (during calibration) and
subtracted out. The residual 2-second spectra were then averaged
together and Hanning-smoothed to obtain the final spectrum for each
epoch. A similar procedure was followed to obtain the final spectrum
towards PKS~B0316+162, whose flux density was measured to be $9.4 \pm
0.7$~Jy in the October session. Intermittent low-level RFI was seen 
near the absorption frequencies in all three runs and careful 
data-editing was hence necessary, especially in the January data.

We also obtained a 602-MHz continuum image of TXS~0311+430 with the 
Giant Metrewave Radio Telescope (GMRT) in March~2007, to determine the 
spatial structure of the quasar radio emission and derive an estimate of 
the covering factor. The total on-source time was 1.5~hours, with a 
16~MHz bandwidth centred at a frequency of 602~MHz and sub-divided into 
128~channels.  The standard calibrator 3C48 was used for flux density 
and bandpass calibration.  These data were analysed in classic AIPS, using 
standard procedures (e.g. \citealt{kanekar07}). 

\section{Spectra and results}

\subsection{Lyman-$\alpha$ and metals}
\label{sec:lya_metals}

\begin{figure}
\centering
\epsfig{file=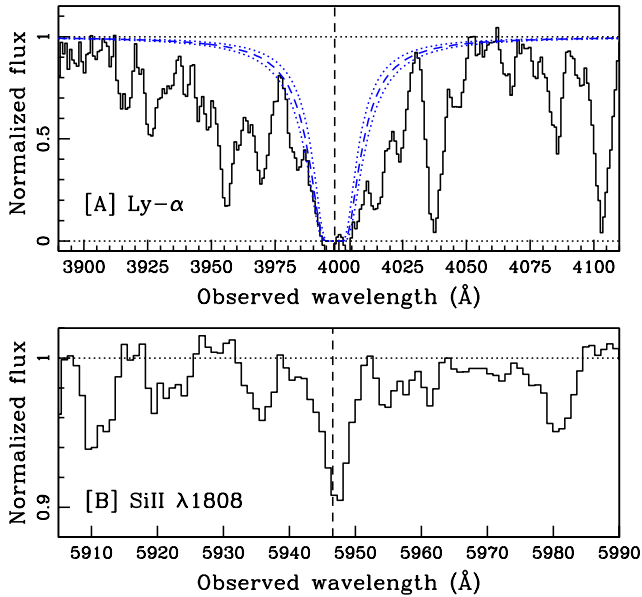,height=3.5truein,width=3.5truein}
\caption{[A]~The damped Lyman-$\alpha$ profile at $z \sim 2.289$ towards
	TXS~0311+430.  The dashed line shows our fit to the profile, which yields
        \nhi\ $= (2.0 \pm 0.5) \times 10^{20}$~\cm. The dotted lines show the 
	$1\sigma$ errors on the fit. The dashed vertical line
        here (and in the lower panel) indicates $z = 2.289$. [B]~The \sii~$\lambda$1808
	line from the $z \sim 2.289$ DLA. }
\label{fig:dla}
\end{figure}

Damped Lyman-$\alpha$ absorption is clearly visible in the GMOS
spectrum towards TXS~0311+430, at $z = 2.289 \pm 0.002$. The 
Lyman-$\alpha$ profile, shown in Fig.~\ref{fig:dla}[A], yields an
\hi\ column density of \nhi\ $= (2.0 \pm 0.5) \times 10^{20}$~\cm, 
derived by overlaying damped profiles using the Starlink {\sc dipso} 
software.  We quote a conservative error of 25\%, to encompass the range 
of reasonable `by-eye' profiles as well as systematic errors from the 
continuum fit. Next, although our spectral coverage was such that only 
a few metal lines associated with the DLA are outside the Lyman-$\alpha$
forest, we were able to detect the \sii~$\lambda$1526,
\sii~$\lambda$1808 and Al~{\sc ii}~$\lambda$1670 transitions at the
DLA redshift; their equivalent widths are listed in Table \ref{tab:metals} 
and the \sii~$\lambda$1808 profile shown in Fig.~\ref{fig:dla}[B]. Despite 
the low resolution of our spectrum ($\sim 200$~\kms), all three 
metal lines have resolved structure, indicating a large velocity width 
and multiple spectral components. Both the Al~{\sc ii}~$\lambda$1670 and 
\sii~$\lambda$1526 lines are usually very strong and heavily saturated in DLAs 
and the high rest frame equivalent widths indicate that this is indeed the case 
for the absorber towards 
TXS~0311+430. Conversely, \sii~$\lambda$1808 is often unsaturated in DLA 
spectra and can therefore be used to derive an abundance for silicon.  However, 
in the present case, the high rest frame equivalent width of the
\sii~$\lambda$1808 line suggests that it too is likely to be
saturated.  Nonetheless, by assuming that the \sii~$\lambda$1808 line
is in the linear part of the curve of growth, it is possible to
obtain a lower limit on the \sii\ column density and hence, on the
metallicity of the DLA. The Si\,{\sc ii} column density is 
$N$(Si\,{\sc ii})\,$\geq 2.7 \times 10^{15}$\,cm$^{-2}$, giving 
[Si/H]\,$\geq -0.48$. This is an unusually high metallicity among DLAs, 
including those in radio-selected QSOs \citep{akerman05}. Among the 104 
DLAs at $z_{\rm abs} > 1.6$ in the compilation by \citet{prochaska07},
there are only three absorbers with higher values of the silicon abundance.

Our estimates of $N$(Si\,{\sc ii}), and hence [Si/H], may be lower limits, 
because we have assumed no line saturation and no depletion of Si onto dust 
grains. On the other hand, it is possible that we may have over-estimated the 
equivalent width of the Si\,{\sc ii}~$\lambda 1808$ absorption line through 
unrecognised blending with other features in our low resolution GMOS spectrum. 
The work of \citet{herbertfort06} ---see, in particular, their Figure~5--- shows 
that in low resolution spectra from the Sloan Digital Sky Survey this effect 
can lead one to overestimate $N$(Si\,{\sc ii}) by as much as a factor of three.
Higher resolution spectra, and observations of other spectral lines such as
Zn\,{\sc ii}~$\lambda\lambda 2026,2062$, should help resolve these ambiguities.

We also examined the GMOS spectrum for possible absorption at other redshifts 
and detected a strong \mgii\ system at $z \sim 1.069$. The large rest frame 
equivalent widths of the \mgii~$\lambda\lambda$2796,2803 doublet ($> 3$~\AA) 
and the \feii~$\lambda$2600 line ($\sim 2.3$~\AA) in this absorber (see 
Table~\ref{tab:metals}) imply that it too is likely to be a DLA \citep{rao06}.

\begin{table}
\begin{center}
\caption{Detected absorption lines in the GMOS spectrum of TXS~0311+430.
All equivalent widths are in the rest frame; errors are 1-$\sigma$, but
do not include systematics due to continuum placement ($<$5\%). 
We do not list an equivalent width for the Mg~{\sc i}~$\lambda$2852 transition 
from the $z \sim 1.069$ absorber, as it appears to be blended with Galactic 
and night-sky Na~{\sc i}~$\lambda$5890 features. }
\begin{tabular}{ccc}
\hline
Transition &    $z$ & EW (m\AA) \\
\hline
\sii\ $\lambda$1526 &           $2.290 \pm 0.003$ &     $761 \pm 58$ \\ 
Al~{\sc ii} $\lambda$1670 &     $2.290 \pm 0.002$ &     $787 \pm 45$ \\ 
\sii\ $\lambda$1808 &           $2.289 \pm 0.002$ &     $169 \pm 32$ \\ 
\hline
\feii\ $\lambda$2344 &          $1.069 \pm 0.002$ &     $2606 \pm 80$ \\
\feii\ $\lambda$2374 &          $1.069 \pm 0.002$ &     $1385 \pm 98$ \\
\feii\ $\lambda$2382 &          $1.069 \pm 0.002$ &     $2374 \pm 84$ \\
\feii\ $\lambda$2586 &          $1.069 \pm 0.002$ &     $1595 \pm 68$ \\
\feii\ $\lambda$2600 &          $1.068 \pm 0.002$ &     $2313 \pm 61$ \\
\mgii\ $\lambda$2796 &          $1.068 \pm 0.001$ &     $3115 \pm 70$ \\
\mgii\ $\lambda$2803 &          $1.069 \pm 0.001$ &     $3109 \pm 72$ \\
Mg~{\sc i}~$\lambda$2852  &     $1.069 \pm 0.001$ &     $-$ \\
\hline
\end{tabular}\label{tab:metals}
\end{center}
\end{table}%

\subsection{Radio spectroscopy and imaging}
\label{sec:radio_results}

\begin{figure}
\centering
\epsfig{file=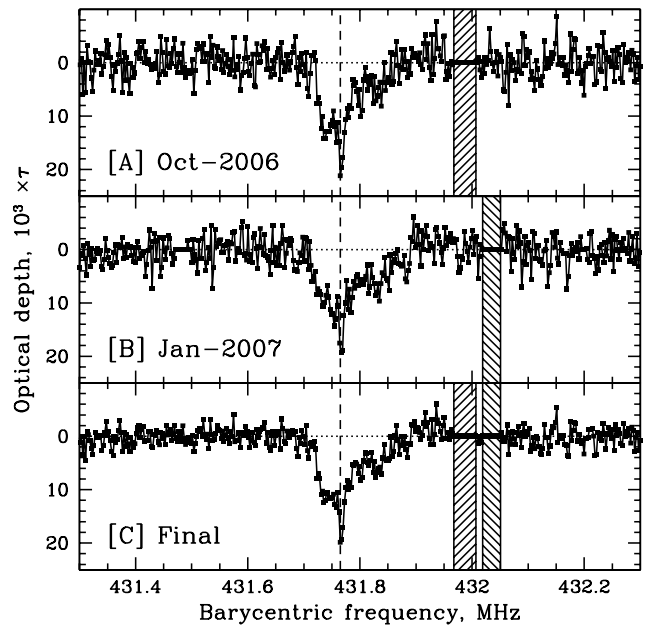,height=3.5truein,width=3.5truein}
\caption{GBT \hi~21~cm spectra towards TXS~0311+430, with optical
depth (in units of $10^3 \times \tau$) plotted against barycentric
frequency, in MHz. The dashed vertical line is at $z =
2.289766$. Panels~[A] and [B] show the spectra from the October and
January runs while panel~[C] shows the final GBT spectrum, obtained by
combining the spectra of panels [A] and [B]. The shaded regions in the
top two panels show frequency ranges affected by RFI; both these
ranges have been blanked out in panel~[C]. }
\label{fig:21cm}
\end{figure}

The final \hi~21~cm spectra from the October and January observing sessions are shown 
in panels~[A] and [B] of Figure~\ref{fig:21cm}, with the vertical shaded region 
in each panel marking a frequency range that was affected by regular RFI in both runs. 
The RMS noise values are $\sim 15.1$~mJy and $\sim 16.5$~mJy per 1.7~\kms\ channel in 
October and January, respectively. A strong absorption feature is visible in both 
spectra between $\sim 431.7$ and $\sim 431.86$~MHz. No spectral features were seen 
in this frequency range in the October spectrum towards the calibrator PKS~B0316+162 (not
shown here), which has an RMS noise of $\sim 30.0$~mJy per 1.7~\kms\ channel. The 
absorption towards TXS~0311+430 also showed the expected doppler shift ($\sim 50$~kHz) 
between the October and January runs.  The lack of any absorption towards
the calibrator source, combined with the expected doppler shift over a 3-month
period, rule out RFI as a possible cause for the absorption seen towards TXS~0311+430.

Figure~\ref{fig:21cm}[C] shows the final GBT 21cm spectrum towards TXS~0311+430,
obtained by averaging the spectra from October and January with appropriate weights 
(based on the RMS noise values), after scaling the January spectrum to a flux density 
of $6.2$~Jy. Regions affected by RFI in either spectrum have been blanked out (note
that these are well-removed from the absorption feature and thus do not affect 
our optical depth measurement). The RMS noise on this spectrum is $\sim 10.4$~mJy
per $\sim 1.7$~\kms\ channel. The 21~cm absorption is complex, extending over $\sim 110$~\kms\ 
(between nulls), with peak opacity at 431.765~MHz, i.e. at $z = 2.289766 (19)$. 
The integrated optical depth is $\int \tau_{21} {\rm dV} = (0.818 \pm 0.085)$~\kms, 
with the error dominated by the uncertainty in the source flux density.

\begin{figure}
\centering
\epsfig{file=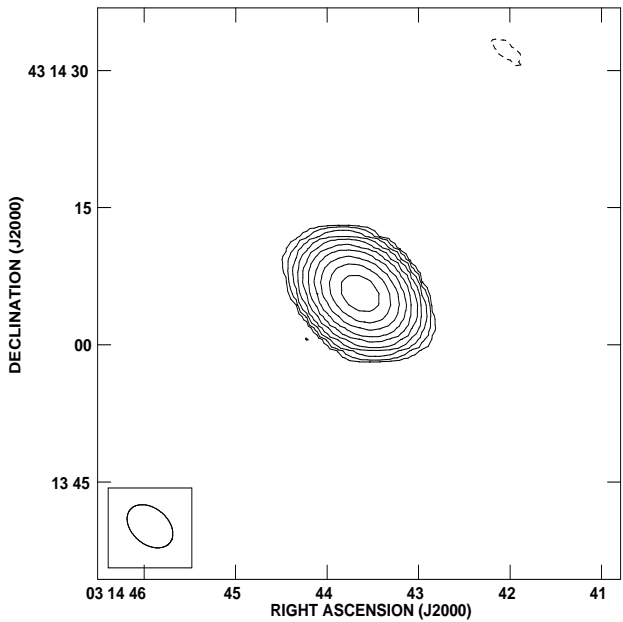,height=3.2truein,width=3.2truein}
\caption{GMRT 602-MHz continuum image of TXS~0311+430, with a resolution of 
$6\farcs0 \times 4\farcs1$ and an RMS noise of 0.6~mJy/Bm. The contours
are at $4.0 \times (-1, 1, 2, 4, 8, 16, 32, 64, 128, 256, 512, 1024)$~mJy.}
\label{fig:im610}
\end{figure}

Figure~\ref{fig:im610} shows the GMRT 602~MHz continuum image of
TXS~0311+430.  The image has a resolution of $6\farcs0 \times
4\farcs1$ (i.e. a spatial resolution of $\sim 50 \times 34 \: h_{71}^{-2}$~kpc$^2$ at
$z = 2.289$\footnote{We use the standard $\Lambda$CDM cosmology, with
$\Omega_m = 0.27$, $\Omega_\Lambda = 0.73$ and H$_0 =
71$~\kms~Mpc$^{-1}$.}) and an RMS noise of 0.6~mJy/Bm; no evidence for
extended structure can be seen. A single elliptical Gaussian component
provides a good fit, yielding an integrated flux density of $3.58$~Jy and a
deconvolved angular size of $\sim 1\farcs36 \times 0\farcs63$; this
gives a spatial extent of $\lesssim 11.3 \times 5.2 \: h_{71}^{-2}$~kpc$^2$ at $z =
2.289$. Of course, this estimate of the source size should be treated
as an upper limit, because any residual phase errors will increase the
observed size. Further, while the continuum image is at a somewhat
different frequency from the redshifted 21~cm line, the lack of any
detected extended emission suggests that the quasar is also very
compact at the latter frequency.

\subsection{The Spin Temperature}
\label{sec:tspin}

For 21~cm absorption studies of DLAs towards radio-loud QSOs, the 21~cm optical depth, 
\hi\ column density \nhi\ and spin temperature $\ts$ are related by the expression 
\begin{equation}
\label{eqn:tspin}
N(HI) = 1.823 \times 10^{18} \left[ \frac{\ts}{f} \right] \intop \tau {\rm d}V,
\end{equation}
where \nhi\ is in cm$^{-2}$, $\ts$ in K, d$V$ in km~s$^{-1}$ and the
21~cm absorption is assumed to be optically thin. In the case of
multiple \hi\ clouds along the line of sight, $\ts$ is the
column-density-weighted harmonic mean of the spin temperatures of the
individual clouds. Note that the use of the above equation to estimate
the spin temperature of the DLA implicitly assumes that the \hi\
column densities along the optical and the (usually more extended)
radio lines of sight are the same.  The covering factor $f$ gives the
fraction of the radio flux density covered by the foreground DLA; this
can be estimated by very long baseline interferometric (VLBI)
observations at or near the redshifted 21cm line frequency, to measure
the fraction of flux density arising from the compact radio core, as
well as its spatial extent (e.g.  \citealt{kanekar07}).

The $z \sim 2.289$ DLA towards TXS~0311+430 has an \hi\ column density
\nhi\ $= (2 \pm 0.5) \times 10^{20}$~\cm\ and an integrated 21~cm
optical depth of $\int \tau {\rm d}V = (0.818 \pm 0.085)$~km/s. No
VLBI measurements are available in the literature at any radio
frequency. However, the sub-arcsecond size estimate from the
GMRT 602~MHz image indicates that the quasar is strongly
core-dominated and suggests a high covering factor, $f \sim 1$. Using
this, we obtain $\ts = (138 \pm 36) \times f$~K, one of the lowest
spin temperatures ever found in a DLA and the only low value currently
known at $z>1$.

\section{Discussion}
\label{sec:discuss}

Spin temperature estimates have so far been obtained in more than
thirty DLAs at all redshifts, including fifteen detections of 21~cm
absorption (Kanekar et al, in preparation).  Only four absorbers, all
at $z < 0.6$, show low spin temperatures, $\ts < 350$~K, typical of
lines of sight through the Milky Way and local disks
\citep{kanekar03}\footnote{Note that this does not include the $z \sim
0.656$ DLA towards 3C336 \citep{curran07}, where the background radio
source is strongly lobe-dominated, with a very small core
fraction. The very extended radio structure (corresponding to a
spatial extent of $\sim 195 \: h_{71}^{-1}$~kpc at $z \sim 0.656$)
implies that the radio and optical absorption almost certainly arise
from very different lines of sight (a possibility discussed by Curran
et al.).}  Another eleven systems, all with
$\ts > 500$~K, have been detected in 21~cm absorption (of which only
four are at $z \gtrsim 2$;
\citealt{wolfe79,wolfe81,kanekar06,kanekar07}) while the remaining
DLAs without detectable 21~cm absorption all have $3\sigma$ lower
limits of $> 700$~K on the spin temperature.

The preponderance of high spin temperature estimates in high-$z$ DLAs
has usually been attributed to a relatively low fraction of the cold
neutral phase of \hi\ (the CNM), with most of the gas in the warm
phase (the WNM) (e.g. \citealt{carilli96,chengalur00}). This assumes
that the absorbers have a similar two-phase structure to that seen in
the ISM of the Milky Way (e.g. \citealt{wolfe03b}).  The low spin temperature of
the $z \sim 2.289$ DLA towards TXS~0311+430 would then imply that a
higher fraction of the \hi\ along the line of sight is in the CNM
compared with the majority of DLAs at $z \gtrsim 2$.  \citet{chengalur00}
argued that the high derived $\ts$ values of high-$z$ DLAs could be
due to the low metallicities of typical high-$z$ DLAs, with the
paucity of metals resulting in fewer radiation pathways for gas
cooling (see also \citealt{young97}). If this is correct, one would
expect DLAs with low spin temperatures (such as the absorber towards
TXS~0311+430) to have significantly higher metallicities than those of
the general DLA population \citep{kanekar01a}. As expected, and assuming 
that the Si\,{\sc ii}~$\lambda 1808$ equivalent width has not been 
over-estimated, the $z \sim 2.289$ DLA has [Si/H]~$\ge -0.48$, one 
of most metal-rich DLAs yet discovered.

It has also been suggested that the observed low 21~cm optical depth
in high-$z$ DLAs arises due to covering factor effects
(e.g. \citealt{curran05}).  In the present case, any reduction in the
covering factor below the assumed value of unity can only strengthen
the case for a low spin temperature (since $\ts = (138 \pm 36) \times
f$~K). The only way to alter the conclusion that a sizeable fraction
of \hi\ along the line of sight is in the cold phase is if there are
large spatial differences in the \hi\ column densities along the
optical and radio lines of sight (e.g. \citealt{wolfe03b}). For
example, if the optical QSO lies behind a ``hole'' in the \hi\ column
density distribution, the average \hi\ column density against the
radio QSO could be larger than that measured from the Lyman-$\alpha$
line, implying a higher spin temperature from
Eqn.~\ref{eqn:tspin}. Unfortunately, it is very difficult to directly
test this possibility.  While Galactic \hi\ column densities derived
from Lyman-$\alpha$ absorption studies are in excellent agreement with
those obtained from \hi\ 21~cm emission observations in the same
directions, despite the very different spatial resolutions in the two
methods, such comparisons have only been carried out for a fairly
small number of high latitude lines of sight \citep{dickey90}.  A
similar comparison between \nhi\ values derived from Lyman-$\alpha$
absorption and 21~cm emission has only been possible in one DLA, the
$z \sim 0.009$ absorber towards SBS~1543+593.  Here, the \hi\ column
densities agree to within a factor of $\sim 2$, despite the extremely
poor spatial resolution ($\sim 5.3 \: h_{71}^{-1}$~kpc) of the radio observations
\citep{chengalur02}. While both of these studies suggest that the
\nhi\ values along the radio and optical lines of sight are likely to
be comparable, we cannot formally rule out the possibility of
differences in an individual absorber. However, the fact that the
expected high metallicity is indeed seen in the $z \sim 2.289$ DLA 
(assuming that our Si~{\sc ii}~$\lambda$1808 measurement is accurate) 
is consistent with the interpretation of a high CNM fraction.

Finally, \citet{kanekar03} noted a relationship between spin
temperature and absorber morphology, in that, at low redshifts, low
$\ts$ values are only found in DLAs identified with luminous disk
galaxies, while low-$z$, low-luminosity DLAs, associated with dwarf or
LSB galaxies, are all found to have high spin temperatures ($\ts
\gtrsim 700$~K).  It has not been possible to test this empirical
relationship at $z > 1$, as the host galaxies of high-$z$ DLAs are
rarely detectable. However, if the relationship does extend out to
high redshifts, we would expect the $z \sim 2.289$ DLA to be a
massive, luminous disk galaxy.  We note that both the high metallicity
and the large velocity spreads seen in the optical low-ionization
metal lines and the 21cm absorption are consistent with the absorption
arising in a massive galaxy. If so, it should be possible to
detect the absorber host with deep imaging; this would be the first direct
test of the $\ts$-morphology relationship at high redshifts.

In summary, we have detected damped Lyman-$\alpha$ and \hi~21~cm absorption at 
$z = 2.289$ towards the quasar TXS~0311+430. We obtain a DLA spin temperature of
$\ts = (138 \pm 36) \times f$~K, the first case of a low spin temperature estimate
in a high redshift DLA. The low spin temperature, high metallicity and large velocity
spread of the 21~cm and metal lines all suggest that the absorber is likely to be a 
massive disk galaxy.

\section{Acknowledgments}
We thank Carl Bignell and Bob Garwood for much help with the GBT observations and 
AIPS++ data analysis. The NRAO is operated by Associated Universities, Inc., 
under cooperative agreement with the NSF. We thank the staff of the GMRT who 
made these observations possible; the GMRT is run by NCRA-TIFR. Based on 
observations obtained at the Gemini Observatory (program ID GN-2005B-Q-60), 
which is operated by the AURA, Inc., under a cooperative agreement with the NSF on 
behalf of the Gemini partnership: the NSF (United States), the STFC (United Kingdom), 
the NRC (Canada), CONICYT (Chile), the ARC (Australia), CNPq (Brazil) and CONICET 
(Argentina). BY was supported in part by a grant from the NSERC (Canada).

\bibliographystyle{mn2e}
\bibliography{ms}

\begin{thebibliography}{}

\bibitem[\protect\citeauthoryear{{Akerman}, {Ellison}, {Pettini} \&
  {Steidel}}{{Akerman} et~al.}{2005}]{akerman05}
{Akerman} C.~J.,  {Ellison} S.~L.,  {Pettini} M.,    {Steidel} C.~C.,  2005,
  A\&A, 440, 499

\bibitem[\protect\citeauthoryear{Braun \& Walterbos}{Braun \&
  Walterbos}{1992}]{braun92}
Braun R.,  Walterbos R.,  1992, ApJ, 386, 120

\bibitem[\protect\citeauthoryear{Briggs \& Wolfe}{Briggs \&
  Wolfe}{1983}]{briggs83}
Briggs F.~H.,  Wolfe A.~M.,  1983, ApJ, 268, 76

\bibitem[\protect\citeauthoryear{Carilli, Lane, {de Bruyn}, Braun \&
  Miley}{Carilli et~al.}{1996}]{carilli96}
Carilli C.~L.,  Lane W.~M.,  {de Bruyn} A.~G.,  Braun R.,    Miley G.~K.,
  1996, AJ, 111, 1830

\bibitem[\protect\citeauthoryear{{Chen} \& {Lanzetta}}{{Chen} \&
  {Lanzetta}}{2003}]{chen03}
{Chen} H.-W.,  {Lanzetta} K.~M.,  2003, ApJ, 597, 706

\bibitem[\protect\citeauthoryear{Chengalur \& Kanekar}{Chengalur \&
  Kanekar}{2000}]{chengalur00}
Chengalur J.~N.,  Kanekar N.,  2000, MNRAS, 318, 303

\bibitem[\protect\citeauthoryear{Chengalur \& Kanekar}{Chengalur \&
  Kanekar}{2002}]{chengalur02}
Chengalur J.~N.,  Kanekar N.,  2002, A\&A, 388, 383

\bibitem[\protect\citeauthoryear{{Curran}, {Murphy}, {Pihlstr{\"o}m}, {Webb} \&
  {Purcell}}{{Curran} et~al.}{2005}]{curran05}
{Curran} S.~J.,  {Murphy} M.~T.,  {Pihlstr{\"o}m} Y.~M.,  {Webb} J.~K.,
  {Purcell} C.~R.,  2005, MNRAS, 356, 1509

\bibitem[\protect\citeauthoryear{{Curran}, {Tzanavaris}, {Murphy}, {Webb} \&
  {Pihlstroem}}{{Curran} et~al.}{2007}]{curran07}
{Curran} S.~J.,  {Tzanavaris} P.,  {Murphy} M.~T.,  {Webb} J.~K.,
  {Pihlstroem} Y.~M.,  2007, MNRAS, in press (astro-ph/0706.2692), 706

\bibitem[\protect\citeauthoryear{{Dickey} \& {Lockman}}{{Dickey} \&
  {Lockman}}{1990}]{dickey90}
{Dickey} J.~M.,  {Lockman} F.~J.,  1990, ARA\&A, 28, 215

\bibitem[\protect\citeauthoryear{{Douglas}, {Bash}, {Bozyan}, {Torrence} \&
  {Wolfe}}{{Douglas} et~al.}{1996}]{douglas96}
{Douglas} J.~N.,  {Bash} F.~N.,  {Bozyan} F.~A.,  {Torrence} G.~W.,    {Wolfe}
  C.,  1996, AJ, 111, 1945

\bibitem[\protect\citeauthoryear{{Ficarra}, {Grueff} \& {Tomassetti}}{{Ficarra}
  et~al.}{1985}]{ficarra85}
{Ficarra} A.,  {Grueff} G.,    {Tomassetti} G.,  1985, A\&AS, 59, 255

\bibitem[\protect\citeauthoryear{{Herbert-Fort}, {Prochaska},
  {Dessauges-Zavadsky}, {Ellison}, {Howk}, {Wolfe} \&
  {Prochter}}{{Herbert-Fort} et~al.}{2006}]{herbertfort06}
{Herbert-Fort} S.,  {Prochaska} J.~X.,  {Dessauges-Zavadsky} M.,  {Ellison}
  S.~L.,  {Howk} J.~C.,  {Wolfe} A.~M.,    {Prochter} G.~E.,  2006, PASP, 118,
  1077

\bibitem[\protect\citeauthoryear{Kanekar \& Chengalur}{Kanekar \&
  Chengalur}{2001}]{kanekar01a}
Kanekar N.,  Chengalur J.~N.,  2001, A\&A, 369, 42

\bibitem[\protect\citeauthoryear{Kanekar \& Chengalur}{Kanekar \&
  Chengalur}{2003}]{kanekar03}
Kanekar N.,  Chengalur J.~N.,  2003, A\&A, 399, 857

\bibitem[\protect\citeauthoryear{Kanekar, Chengalur \& Lane}{Kanekar
  et~al.}{2007}]{kanekar07}
Kanekar N.,  Chengalur J.~N.,    Lane W.~M.,  2007, MNRAS, 375, 1528

\bibitem[\protect\citeauthoryear{Kanekar, Subrahmanyan, Ellison, Lane \&
  Chengalur}{Kanekar et~al.}{2006}]{kanekar06}
Kanekar N.,  Subrahmanyan R.,  Ellison S.~L.,  Lane W.~M.,    Chengalur J.~N.,
  2006, MNRAS, 370, L46

\bibitem[\protect\citeauthoryear{{Kulkarni}, {Fall}, {Lauroesch}, {York},
  {Welty}, {Khare} \& {Truran}}{{Kulkarni} et~al.}{2005}]{kulkarni05}
{Kulkarni} V.~P.,  {Fall} S.~M.,  {Lauroesch} J.~T.,  {York} D.~G.,  {Welty}
  D.~E.,  {Khare} P.,    {Truran} J.~W.,  2005, ApJ, 618, 68

\bibitem[\protect\citeauthoryear{{Lodders}}{{Lodders}}{2003}]{lodders03}
{Lodders} K.,  2003, ApJ, 591, 1220

\bibitem[\protect\citeauthoryear{Prochaska, Wolfe, Howk, Gawiser, Burles \&
  Cooke}{Prochaska et~al.}{2007}]{prochaska07}
Prochaska J.~X.,  Wolfe A.~M.,  Howk J.~C.,  Gawiser E.,  Burles S.~M.,
  Cooke J.,  2007, ApJS, 171, 29

\bibitem[\protect\citeauthoryear{{Rao}, {Turnshek} \& {Nestor}}{{Rao}
  et~al.}{2006}]{rao06}
{Rao} S.~M.,  {Turnshek} D.~A.,    {Nestor} D.~B.,  2006, ApJ, 636, 610

\bibitem[\protect\citeauthoryear{{Wolfe} \& {Briggs}}{{Wolfe} \&
  {Briggs}}{1981}]{wolfe81}
{Wolfe} A.~M.,  {Briggs} F.~H.,  1981, ApJ, 248, 460

\bibitem[\protect\citeauthoryear{Wolfe \& Davis}{Wolfe \&
  Davis}{1979}]{wolfe79}
Wolfe A.~M.,  Davis M.~M.,  1979, AJ, 84, 699

\bibitem[\protect\citeauthoryear{Wolfe, Gawiser \& Prochaska}{Wolfe
  et~al.}{2003}]{wolfe03b}
Wolfe A.~M.,  Gawiser E.,    Prochaska J.~X.,  2003, ApJ, 593, 235

\bibitem[\protect\citeauthoryear{Wolfe, Gawiser \& Prochaska}{Wolfe
  et~al.}{2005}]{wolfe05}
Wolfe A.~M.,  Gawiser E.,    Prochaska J.~X.,  2005, ARA\&A, 43, 861

\bibitem[\protect\citeauthoryear{{Young} \& {Lo}}{{Young} \&
  {Lo}}{1997}]{young97}
{Young} L.~M.,  {Lo} K.~Y.,  1997, ApJ, 490, 710

\end{thebibliography}

\label{lastpage}
\end{document}